\documentclass[12pt]{iopart}

\usepackage{bm}
\usepackage{dcolumn}
\usepackage{longtable}
\usepackage{multirow}
\usepackage{graphicx}

\newcommand{\lambdabar}{\lambda\kern-.5em\raise.5ex\hbox{--}}
\newcommand{\Ninej}[9]{\left\{\matrix{
 {#1} & {\!\!\!#2} & {\!\!\!#3}\cr
 {#4} & {\!\!\!#5} & {\!\!\!#6}\cr
 {#7} & {\!\!\!#8} & {\!\!\!#9}\cr}\right\}}
\newcommand{\Cleb}[6]{C^{{\,#1}{\,#2}{\,#3}}
  _{{\,#4}{\,#5}{\,#6}} }

\begin{document}

\title[Bound states of He(2 $^{3}$S)+He(2 $^{3}$P)]
{Effects of non-adiabatic and Coriolis couplings on the bound states of He(2 $^{3}$S)+He(2 $^{3}$P)}

\author{D G Cocks$^1$, I B Whittingham$^1$ and G Peach$^2$}

\address{$^1$ School of Engineering and Physical Sciences, 
James Cook University, Townsville 4811, Australia} 

\address{$^2$ Department of Physics and Astronomy, University College
London, London WC1E~6BT, UK}

\ead{daniel.cocks@jcu.edu.au}

\begin{abstract}
The effects of non-adiabatic and Coriolis couplings on the bound states of the 
\mbox{He(2 $^{3}$S$_{1}$)+He(2 $^{3}$P$_{j}$)} system, where $j=0,1,2$, are 
investigated using the recently available  \textit{ab initio} short-range 
${}^{1,3,5}\Sigma^{+}_{g,u}$ and ${}^{1,3,5}\Pi_{g,u}$ potentials computed by 
Deguilhem \textit{et al.} (\textit{J. Phys. B: At. Mol. Opt. Phys.} \textbf{42} 
(2009) 015102).  Three sets of calculations have been undertaken: 
single-channel, multichannel without Coriolis couplings and full multichannel 
with Coriolis couplings.  We find that non-adiabatic effects are negligible for 
$0^{-}_{u},0^{\pm}_{g},1_{u},2_{g},2_{u},3_{g}$ Hund case (c) sets of levels in 
the $j=2$ asymptote but can be up to 15\% for some of the $0^{+}_{u}$ and 
$1_{g}$ sets of levels where near degeneracies are present in the single-channel 
diagonalized potentials.  Coriolis couplings are most significant for weakly 
bound levels, ranging from 1-5\% for total angular momenta $J=1,2$ and up to 
10\% for $J=3$. Levels near the $j=1$ and $j=0$ asymptotes agree closely with 
previous multichannel calculations based upon long-range potentials constructed 
from retarded resonance dipole and dispersion interactions.  Assignment of 
theoretical levels to experimental observations using criteria based upon the 
short-range character of each level and their coupling to metastable ground 
states produces well matched assignments for the majority of observations. After 
a 1\% increase in the slope of the ${}^5\Sigma^+_{g,u}$ and ${}^5\Pi_{g,u}$ 
input potentials near the classical turning point is applied, improved matching 
of previous assignments is obtained and further assignments can be made, 
reproducing very closely the number of experimental observations.

\end{abstract}


\submitto{\jpb}
\maketitle

\section{Introduction}

Photoassociation (PA) of ultracold atoms provides a powerful technique for the 
study of the dynamics of ultracold collisions. The two interacting ultracold 
atoms are resonantly excited by a laser to bound states of the associated 
molecule with the transition energies forming a spectrum with a very high 
resolution of $< 1$ MHz, since the thermal distribution of energies in the 
initial scattering state is very narrow.

Photoassociation in metastable rare gases is of particular interest as the 
large
internal energy can be released during collisions and provide experimental 
strategies for the study of these quantum gases. A number of experimental 
investigations have been conducted using PA in metastable helium as the 
diagnostic tool.  Bound states that dissociate to the 
2s$\,{}^{3}$S$_{1}+$2p$\,{}^{3}$P$_{2}$ limit were first observed by Herschbach 
\textit{et al.} \cite{Hersch00} and more recently, Kim \textit{et al.} 
\cite{Kim04} and van Rijnbach \cite{Rijn04} have observed detailed structure of 
over 40 peaks associated with bound states with binding energies $\leq $ 13.57 
GHz that dissociate to this limit. In addition, van Rijnbach \cite{Rijn04} has 
observed six peaks lying within 0.6 GHz of the 
2s$\,{}^{3}$S$_{1}+$2p$\,{}^{3}$P$_{1}$ limit and L\'{e}onard \textit{et al.} 
\cite{Leonard03} have studied some purely long-range bound states with binding 
energies $\leq $ 1.43 GHz dissociating to the 
2s$\,{}^{3}$S$_{1}+$2p$\,{}^{3}$P$_{0}$ limit.  

The bound states dissociating to the 2s$\,{}^{3}$S$_{1}+$2p$\,{}^{3}$P$_{0}$
limit occur at interatomic separations $\geq 150\,a_{0}$ and arise from 
resonance dipole and dispersion interactions that depend upon well-known atomic 
parameters.  Theoretical analyses have been completed using both a 
single-channel adiabatic calculation \cite{Leonard04} and full multichannel 
calculations \cite{Venturi03} that employ long-range Born-Oppenheimer potentials 
constructed from retarded resonance dipole and dispersion interactions.  
Excellent agreement is obtained with the measured binding energies.  The 
numerous observed peaks associated with the 2s$\,{}^{3}$S$_{1}+$2p$\,{}^{3}$P$_{1,2}$   
limits are not due to long-range states. Most of the peaks were identified by 
\cite{DGL05} using the accumulated phase technique for a single-channel 
calculation of the bound states based upon a hybrid quintet potential 
constructed from short-range \textit{ab initio} ${}^{5}\Sigma^{+}_{g,u}$ and 
${}^{5}\Pi^{+}_{g,u}$ potentials matched onto long-range retarded resonance 
dipole and dispersion potentials.

Recently Deguilhem \etal \cite{DLGD09} have reanalyzed the PA peaks associated 
with the 2s$\,{}^{3}$S$_{1}+$2p$\,{}^{3}$P$_{1,2}$ limits using new fully 
\textit{ab initio} multi-configuration self-consistent field (MCSCF) short-range 
${}^{1,3}\Sigma^{+}_{g,u}$ and ${}^{1,3}\Pi_{g,u}$ potentials and 
multi-reference configuration interaction (MRCI) ${}^{5}\Sigma^{+}_{g,u}$ and 
${}^{5}\Pi_{g,u}$ potentials.  The body-fixed Hamiltonian in the Hund case (c) 
basis is diagonalized and the resulting adiabatic potentials used in a 
single-channel calculation, thus neglecting Coriolis and non-adiabatic 
couplings.  Although earlier multichannel calculations for the ultra-long-range 
states \cite{Venturi03} show this approximation to be quite accurate for these 
states, the effects of these neglected couplings on the numerous more 
strongly-bound shorter-range states warrants investigation. With the 
availability of the short-range ${}^{1,3,5}\Sigma^{+}_{g,u}$ and 
${}^{1,3,5}\Pi_{g,u}$ potentials it is now possible to extend these multichannel 
calculations to the full set of bound states of the 
2s$\,{}^{3}$S$_{1}+$2p$\,{}^{3}$P$_{0,1,2}$ system.

We report here our results of such calculations, together with an analysis of 
the applicability of single-channel calculations. Atomic units are used, with 
lengths in Bohr radii $a_{0} = 0.0529177209$ nm and energies in Hartree 
$E_{h}=\alpha ^{2}m_{e}c^{2}=27.211384$ eV.

\section{Theory}
\subsection{Coupled-channel approach}

The bound rovibrational levels of the ultracold excited metastable helium system 
are found by analyzing the eigenvalues of the molecular Hamiltonian
\begin{equation}
\label{cwp1}
\hat{H} = \hat{T} + \hat{H}_\mathrm{rot} + \hat{H}_\mathrm{el} +
\hat{H}_\mathrm{fs}
\end{equation}
where $\hat{T}$ is the kinetic energy operator
\begin{equation}
\label{kinetic_operator}
   \hat{T}=-\frac{\hbar ^{2}}{2\mu R^{2}}\frac{\partial}{\partial R}
   \left( R^{2}\frac{\partial}{\partial R}\right)\,,
\end{equation}
and $\hat{H}_{\mathrm{rot}}$ the rotational operator
\begin{equation}
\label{rot_operator}
   \hat{H}_{\mathrm{rot}} = \frac{\hat{l}^{2}}{2 \mu R^{2}}\,,
\end{equation}
for a system of two atoms $i=1,2$ with interatomic separation $R$, reduced 
mass $\mu $ and relative angular momentum $\hat{\bm{l}}$. The total
electronic Hamiltonian is 
\begin{equation}
\label{cwp1a}
\hat{H}_{\mathrm{el}}=\hat{H}_{1}+\hat{H}_{2}+\hat{H}_{12}
\end{equation}
where the unperturbed atoms have Hamiltonians $\hat{H}_{i}$ and their 
electrostatic interaction  is specified by $\hat{H}_{12}$. The term 
$\hat{H}_{\mathrm{fs}}$ in equation (\ref{cwp1}) describes the 
fine structure of the atoms.  

The multichannel equations describing the interacting atoms are obtained from 
the eigenvalue equation
\begin{equation}
\label{cwp2}
\hat{H} |\Psi\rangle = E |\Psi \rangle
\end{equation}
by expanding the eigenvector in terms of a basis of the form $|\Phi_a\rangle = 
|\Phi_a (R,q) \rangle$ where $a$ denotes the set of approximate quantum numbers 
describing the electronic-rotational states of the molecule and $q$ denotes the 
interatomic polar coordinates $(\theta ,\phi)$ and electronic coordinates 
$(\bm{r}_{1},\bm{r}_{2})$. The expansion
\begin{equation}
\label{cwp3}
|\Psi \rangle = \sum_a \frac{1}{R} G_a(R) |\Phi_a\rangle
\end{equation}
yields the multichannel equations
\begin{equation}
\label{cwp4}
\sum_a \left\{ T^{G}_{a^\prime a}(R) + \left[V_{a^\prime a}(R) - E 
\delta_{a^{\prime}a}\right] G_{a}(R)\right\} = 0\,,
\end{equation}
where
\begin{equation}
\label{cwp5}
T^{G}_{a^{\prime} a}(R) = -\frac{\hbar^{2}}{2 \mu }\langle \Phi_{a^{\prime}}| 
\frac{\partial ^{2}}{\partial R^{2}} G_{a}(R)|\Phi_{a} \rangle
\end{equation}
and
\begin{equation}
\label{cwp6}
V_{a^{\prime} a}(R) = \langle \Phi _{a^{\prime}}| \left[\hat{H}_{\mathrm{rot}} + 
\hat{H}_{\mathrm{el}} + \hat{H}_{\mathrm{fs}}\right] |\Phi_{a} \rangle \,.
\end{equation}

For two colliding atoms with orbital $\hat{\bm{L}}_{i}$, spin $\hat{\bm{S}}_{i}$ 
and total $\hat{\bm{j}}_{i}$ angular momenta, several different basis 
representations can be constructed. Two possibilities are the $LS$ coupling 
scheme $\hat{\bm{L}}=\hat{\bm{L}}_{1}+\hat{\bm{L}}_{2}$, 
$\hat{\bm{S}}=\hat{\bm{S}}_{1}+\hat{\bm{S}}_{2}$ and 
$\hat{\bm{J}}=\hat{\bm{L}}+\hat{\bm{S}}+\hat{\bm{l}}$ and the $jj$ coupling 
scheme $\hat{\bm{j}}_{1}=\hat{\bm{L}}_{1}+\hat{\bm{S}}_{1}$, 
$\hat{\bm{j}}_{2}=\hat{\bm{L}}_{2}+\hat{\bm{S}}_{2}$, $\hat{\bm{j}}= 
\hat{\bm{j}}_{1}+\hat{\bm{j}}_{2}$ and $\hat{\bm{J}}=\hat{\bm{j}}+\hat{\bm{l}}$.  
The $LS$ coupling scheme diagonalizes $\hat{H}_{\mathrm{el}}$ whereas the $jj$ 
coupling scheme diagonalizes $\hat{H}_{\mathrm{fs}}$.  We choose to use the 
body-fixed $jj$ coupled states \cite{CW09}
\begin{equation}
\label{cwp7}
|\gamma_{1}\gamma_{2}j_{1}j_{2}j\Omega_{j}wJm_{J}\rangle = 
\sqrt{\frac{2J+1}{4\pi}}
D_{m_{J}\Omega_{J}}^{J*}(\phi,\theta,0)|\gamma_{1} \gamma_{2} 
j_{1}j_{2}j\Omega_{j}w\rangle \,,
\end{equation}
where $\gamma_{i}$ represents other relevant quantum numbers such as 
$\{L_{i},S_{i}\}$.  The projections of $j$ and $J$ respectively onto the 
inter-molecular axis $OZ$ are specified by $\Omega_{j}$ and 
$\Omega_{J}=\Omega_{j}$ which has orientation $(\theta,\phi)$ relative to the 
space-fixed frame.  The symmetry under inversion of the electronic wavefunction 
through the centre of charge is denoted by $w$ which is equal to \textit{gerade} 
$(g)$ or \textit{ungerade} $(u)$.  The projection of $\hat{\bm{J}}$ onto the 
space-fixed $Oz$ axis is labelled by $m_{J}$ and 
$D_{m_{J}\Omega_{J}}^{J*}(\phi,\theta,0)$ is the Wigner rotation matrix 
\cite{BrinkSatchler}.

The matrix elements of the various contributions to the Hamiltonian in this 
basis are derived in \cite{CW09}. We list here the required elements using the 
abbreviated notation $|a \rangle = |\Phi_{a}(R,q) \rangle $ where 
$a=\{\gamma_{1},\gamma_{2},j_{1},j_{2},j,\Omega_{j},w,J,m_{J}\}$.  The radial 
kinetic energy terms are
\begin{equation}
\label{cwp8}
\langle a^{\prime}|\hat{T}\frac{1}{R}G_{a}(R)|a\rangle =
 -\frac{\hbar^{2}}{2\mu R}
\frac{d^{2}G_{a}}{dR^{2}}\delta_{a a^{\prime}}
\end{equation}
if we assume the $R-$dependence of the basis states is negligible.  The rotation 
terms are given by
\begin{eqnarray}
\label{cwp9}
\fl
\langle a^{\prime}|\hat{l^{2}}|a\rangle  =  \hbar^{2} \delta_{\rho 
\rho^{\prime}}\left\{\left[J(J+1)+j(j+1)-2\Omega_{j}^{2}\right]
\delta_{\Omega_{j}^{\prime}\Omega_{j}}
\right. \nonumber \\  \left.  
-\;K^{-}_{Jj\Omega_{j}}\delta_{\Omega_{j}^{\prime},\Omega_{j}-1} -
K^{+}_{Jj\Omega_{j}}\delta_{\Omega_{j}^{\prime},\Omega_{j}+1}
\right\}\,,
\end{eqnarray}
where the quantities
\begin{eqnarray}
\label{cwp10}
\fl
K^{\pm}_{Jj\Omega_{j}} = \left[J(J+1) - \Omega_j(\Omega_j \pm 
1)\right]^{\frac{1}{2}}
\nonumber \\  \times
\left[ j(j+1) - \Omega_j (\Omega_j \pm 1)\right]^{\frac{1}{2}}
\end{eqnarray}
describe the Coriolis couplings and $\rho$ denotes the set of quantum numbers 
$\{\gamma_{1},\gamma_{2},j_{1},j_{2},j,w,J,m_{J}\}$.  The electronic terms are
\begin{eqnarray}
\label{cwp11}
\fl
\langle a^{\prime}|\hat{H}_{\mathrm{el}}|a\rangle  =  \delta_{\eta 
\eta^{\prime}}\sum_{LS\Omega_{L}\Omega_{S}}F_{LS\Omega_{L}
\Omega_{S}}^{j_{1}^{\prime}j_{2}^{\prime}j^{\prime}\Omega_{j}}
\nonumber  \\ \times
\; [{}^{2S+1}\Lambda^{\sigma}_{w}(R)+E_{a}^{\infty}] 
F_{LS\Omega_{L}\Omega_{S}}^{j_{1}j_{2}j\Omega_{j}} \,,
\end{eqnarray}
where the coupling coefficients 
$F_{LS\Omega_{L}\Omega_{S}}^{j_{1}j_{2}j\Omega_{j}}$ are given in 
\ref{app:decomposition}, $\Lambda =|\Omega_{L}|$, 
${}^{2S+1}\Lambda^{\sigma}_{w}(R)$ are the Born-Oppenheimer (BO) molecular 
potentials for \textit{gerade} $(w=0)$ and \textit{ungerade} $(w=1)$ symmetries, 
$\sigma$ is the symmetry of the electronic wave function with respect to 
reflection through a plane containing the internuclear axis and $E^{\infty}_{a}$ 
is the asymptotic energy of the state.  The label $\eta$ denotes the set of 
quantum numbers $\{\gamma_{1},\gamma_{2},\Omega_{j},w,J,m_{J}\}$. Finally, we 
assume that the fine structure is independent of $R$ so that its contribution is
\begin{equation}
\label{cwp13}
\langle a^{\prime}|\hat{H}_{\mathrm{fs}}|a\rangle = \delta_{aa^{\prime}} (\Delta 
E_{\gamma_{1}j_{1}}^{\mathrm{fs}}+\Delta E_{\gamma_{2}j_{2}}^{\mathrm{fs}})\,.
\end{equation}
The fine structure splitting $\Delta E_{\gamma_{1}j_{1}}^{\mathrm{fs}}$ for the 
2s$\, {}^{3}$S$_{1}$ level vanishes and the splittings $\Delta 
E_{\gamma_{2}j_{2}}^{\mathrm{fs}}$ for the 2p$\,{}^3$P$_0$ and 2p$\,{}^3$P$_1$ 
states relative to the 2p$\,{}^3$P$_2$ level are $31.9081$ GHz and $2.2912$ GHz 
respectively.  The matrix elements of the total Hamiltonian are therefore 
diagonal in $\{w,J,m_{J}\}$.

The coupled equations (\ref{cwp4}) then become
\begin{eqnarray}
\label{cwp4a}
\fl
\sum_{a} \left\{ -\left[ \frac{\hbar^{2}}{2 \mu }\frac{\rmd^{2}}{\rmd R^{2}} + E 
- \Delta E_{\gamma_{2}j_{2}}^{\mathrm{fs}}\right]\delta_{a^{\prime}a} + 
\frac{\langle a^{\prime} |\hat{l}^{2} |a \rangle}{2 \mu R^{2}}
\right.  \nonumber  \\ \left.
+ \langle a^{\prime} | \hat{H}_{\mathrm{el}}|a \rangle \right\} G_{a}(R) = 0\,.
\end{eqnarray}
In the present investigation we seek the $m_{J}$-degenerate discrete eigenvalues 
$E=E_{a,v}$ and associated radial eigenfunctions $G_{a,v}(R)$, where $v=0,1, 
\ldots$ labels the rovibrational levels.

In common use throughout the literature of coupled-channel calculations are the 
terms `multichannel' and `close-coupled'. The term `close-coupled' is often used 
to refer to coupled systems for which the states of interest are only strongly 
coupled directly to a small number of neighbouring states but may be indirectly 
coupled, albeit very weakly, to an endless series of states. In this situation, 
a limit must be imposed upon the number of states to be included in the model. 
In contrast, our present full multichannel calculations include all the
coupled 2s2p states and these provide a complete basis except for the negligible 
couplings to other electronic states that may occur at very small interatomic 
distances.

\subsection{Single-channel approximation}

In many situations the Coriolis couplings can be neglected and the calculation 
of the bound states reduced to that for a single channel. This is also useful in 
assigning the appropriate quantum numbers to the states found by the 
multichannel method.  The Movre-Pichler model \cite{MP77}, extended to 
include rotation, is used. At each value of $R$ the single-channel potential 
is formed by diagonalizing the matrix:
\begin{equation}
\label{cwp20}
V_{a\prime a}^{\Omega_{j}} = \langle a^{\prime} | \hat{H}_{\mathrm{el}}|a 
\rangle + \langle a^{\prime}|\hat{H}_{\mathrm{fs}}|a \rangle +\frac{\langle 
a^{\prime} |\hat{l}^{2} |a \rangle_{\Omega_{j}}} {2 \mu R^{2}} \,,
\end{equation}
where $\langle a^{\prime} |\hat{l}^{2} |a \rangle _{\Omega_{j}}$ is the part of 
(\ref{cwp9}) diagonal in $\Omega_{j}$.  The corresponding $R$-dependent 
eigenvectors are
\begin{equation}
\label{cwp21}
|i \rangle = \sum_{a} C_{ai} (R) |a \rangle
\end{equation}
and the adiabatic potential is given by $V^\mathrm{adi}_i(R) = 
\sum_{a^{\prime}a}C^{-1}_{a^{\prime}i}V^{\Omega_j}_{a^{\prime}a}C_{ai}$.  Since 
we assume that the Coriolis couplings are negligible, each channel $|i \rangle $ 
can be labelled with the Hund's case (c) notation $\{J,\Omega_w^\sigma \}$ where 
$\Omega =|\Omega_{J}|$. The adiabatic eigenvalue equation for the 
rovibrational eigenstates $|\psi_{i,v} \rangle = R^{-1}G_{i,v}(R)|i\rangle$, 
where $i=\{J,\Omega_w^\sigma \}$, is then obtained by neglecting the 
off-diagonal elements of the kinetic energy due to non-adiabatic couplings, so 
that
\begin{eqnarray}
\label{cwp22}
\fl
\langle i^{\prime}|\hat{T}\frac{1}{R}G_{i,v}(R)|i \rangle = 
-\frac{\hbar^{2}}{2\mu R} \left(\frac{d^{2}G_{i,v}}{dR^{2}}\delta_{i i^{\prime}}
\right. \nonumber  \\ \left.
+ 2 \frac{\rmd G_{i,v}}{\rmd R} \langle i^{\prime}|\frac{\rmd }{\rmd R} |i 
\rangle + G_{i,v}\langle i^{\prime} |\frac{\rmd ^{2}}{\rmd R^{2}}|i \rangle 
\right) .
\end{eqnarray}
Since the term $\langle i |\rmd |i \rangle /\rmd R = (1/2)\rmd \langle i|i 
\rangle/\rmd R$ vanishes, the radial eigenvalue equation for the rovibrational 
states is
\begin{equation}
\label{cwp23}
\left\{ -\frac{\hbar^2}{2\mu }\left[ \frac{\rmd^2}{\rmd R^2} + 
U^\mathrm{KC}_{i}(R)\right] + V^\mathrm{adi}_{i}(R) - E_{i,v} \right\} 
G_{i,v}(R) = 0\,.
\end{equation}
The kinetic energy correction term,
\begin{equation}
\label{cwp24}
U^\mathrm{KC}_{i}(R)=\langle i|\frac{\rmd ^{2}}{\rmd R^{2}}|i \rangle
= \sum_{a}C_{ai}(R) \frac{\rmd^{2}C_{ai}(R)}{\rmd R^{2}}\,,
\end{equation}
arising from the $R$-dependence of the diagonalized basis is small and 
calculable away from potential crossings but cannot be used near crossings 
because of discontinuities in $C_{ai}(R)$ resulting from the diabatic behaviour 
introduced through the use of a finite diagonalization mesh.  It is not used in 
the present calculations.

\subsection{Input potentials}

The required Born-Oppenheimer potentials ${}^{1,3,5}\Sigma^{+}_{g,u}$ and 
${}^{1,3,5}\Pi_{g,u}$ were constructed by matching the \textit{ab initio} MCSCF 
and MRCI short-range potentials of \cite{Deguil09} onto the long-range 
dipole-dipole plus dispersion potentials given by \cite{Venturi03} so that
\begin{equation}
\label{cwp25}
V_{\Lambda }^{\mathrm{long}}(R)= -f_{3\Lambda}(R/\lambdabar)
C_{3\Lambda}/R^3 - C_{6\Lambda}/R^6 - C_{8\Lambda}^\pm / R^8,
\end{equation}
where $f_{3\Lambda}$ is an $R$- and $\Lambda$-dependent retardation 
correction~\cite{Meath68}. The wavelength for the transition 
2s$\,{}^3$S--2p$\,{}^3$P is $\lambda$, where $\lambdabar=\lambda/(2\pi)=3258.17 
\, a_0$.  The $C_{3\Sigma}$ coefficient is $\pm 2 C_3$ and $C_{3\Pi}$ is $\pm 
C_3$, where $C_3 =6.41022 \, E_h a_0^3$.  The contributions, $C_{3\Lambda}/R^3$, 
to the potentials are attractive (repulsive) for $S+w$ odd (even).
For the van der Waals coefficients we use $C_{6\Sigma}=2620.76 \,E_h a_0^6$ and 
$C_{6\Pi}=1846.60 \, E_h a_0^6$. The $C^\pm_{8\Lambda}$ terms are 
$C^+_{8\Sigma}=1515383 \, E_h a_0^8$, $C^-_{8\Sigma}=297215.9\,E_h a_0^8$, 
$C^+_{8\Pi}=97244.75 \, E_h a_0^8$ and $C^-_{8\Pi}=162763.8\,E_h a_0^8$ 
\cite{Marinescu} where the superscript indicates the sign of $(-1)^{S+w}$.

The matching of the short-range \textit{ab initio} and long-range dipole-dipole 
plus dispersion potentials was undertaken at 30 $a_{0}$ for the singlet and 
triplet potentials and 100 $a_{0}$ for the quintet potentials \cite{Deguil09}.  
At these matching points, $R_{\mathrm{m}}$, the short-range potentials were 
shifted to agree with the long-range potentials. 
Although the derivatives of the short-range and long-range potentials at $R_{m}$ 
were equal to within the accuracy of the potentials, a spline fit to the shifted 
tabulated short-range potentials was performed for $R \leq 
R_{\mathrm{m}}+3\,a_{0}$ using the long-range potentials at the additional 
points $R_{m} < R \leq R_{m}+3\,a_{0}$ to obtain smoother matching at $R_{m}$.  
The analytical form (\ref{cwp25}) for the long-range potentials was used for $R 
> R_{\mathrm{m}}+3\,a_{0}$.  

Although the short-range potentials are tabulated to within the classically 
forbidden region, some small variation of binding energies is possible if the 
$R\rightarrow 0$ extrapolation is modified. To be consistent, an extrapolation 
of the form $1/R + A + BR^2$ is used to emulate the expected behaviour for small 
interatomic distances \cite{Bingel59}. The constants $A$ and $B$ are determined 
from the two innermost tabulated values of the potentials.

\section{Numerical issues}

The coupled-channel equations (\ref{cwp4a}) and the single-channel equation 
(\ref{cwp23}) are of the form
\begin{equation}
\label{cwp30}
\left[ \mathbf{I} \frac{\rmd^{2}}{\rmd R^{2}} + \mathbf{Q}(R)\right] 
\mathbf{G}(R) = 0\,,
\end{equation}
where for the case of coupled-channels, $\mathbf{G}$ is the matrix of solutions 
with the second subscript labelling the linearly independent solutions.  The 
integration region $0 \leq R \leq R_{\infty}$ was divided into a number of 
regions $\mathcal{R}_\alpha $, each containing equally spaced grid points, with 
a step size $\Delta R_{\alpha} =0.0001\,a_{0}$ for the region closest to the origin, 
increasing to $0.2\,a_{0}$ for the outermost region $R>1000\,a_{0}$.  
These equations were solved using the renormalized Numerov method \cite{Johnson78}
with the solutions for neighbouring regions matched using the values of the 
functions and their derivatives on the boundaries of the regions.  
The eigenvalues  were determined by a 
bisection technique based on counting the nodes of the determinant 
$|\mathbf{G}(R)|$.

The adiabatic potentials were computed on an equally spaced diagonalization grid 
$R_{n}$.  For a sufficiently fine grid, crossings in the coupled diabatic 
potentials become avoided crossings in the adiabatic potentials regardless of 
coupling strength.  However, for weak coupling between diabatic potentials, the 
probability that the system follows the diabatic path is much greater than that 
for an adiabatic path.  We choose to emulate this diabaticity by creating hybrid 
adiabatic/diabatic single-channel potentials. These potentials are formed by 
applying the function
\begin{equation}
\label{cwp30a}
Y(R_n) = \frac{2 X(R_n)}{X(R_{n-1}) + X(R_{n+1})}\,,
\end{equation}
where $X(R_n) = |V^\mathrm{adi}_a(R_n) - V^\mathrm{adi}_b(R_n)|$, to each pair 
of adiabatic potentials ${a,b}$. Crossings are located at points $R = R_n$ where 
$Y(R_n) \leq 1$ and are treated as either an avoided crossing for 
$Y(R_n)>\alpha$ or a true crossing for $Y(R_n) \leq \alpha$ by interchanging the 
potentials appropriately, where the parameter $\alpha$ represents the ratio of 
diabatic to adiabatic behaviour at the crossings.  The resultant single-channel 
potentials $V^\mathrm{SC}_k (R_n)$ possess diabatic behaviour when required, yet 
also allow avoided crossings to occur in some crucial regions. The 
single-channel basis states are then
\begin{equation}
\label{cwp30b}
|k\rangle=\sum_{a}D_{ak}(R_{n})|a\rangle\,,
\end{equation}
where $D_{ak}$ is equivalent to $C_{ai}$ except for interchanges of columns that 
correspond to the interchanges of the potentials described above.  The choice of 
diagonalization grid size $\Delta R_n$ significantly affects the formation of 
the single-channel potentials, and the values $\Delta R_n = 0.01$ $a_0$ and 
$\alpha = 0.5$ were chosen as these most closely match the multichannel results.

\section{Results}
\subsection{Calculations}

Three sets of calculations have been undertaken: (i) single-channel (SC) which 
ignores non-adiabatic and Coriolis couplings, (ii) multichannel without 
inclusion of Coriolis couplings (MC1) and (iii) full multichannel that include 
Coriolis couplings (MC2).  The SC levels are labelled by the Hund's case (c) 
notation $\{J,\Omega_w^\sigma \}$, the total electronic angular momentum 
$j=0,1,2$ of the asymptotic 2s$\,{}^{3}$S$_{1}+$2p$\,{}^{3}$P$_{j}$ limit and to 
distinguish any remaining multiplicity, the minima of the potentials.  The MC1 
levels are labelled by  $\{J,\Omega_w^\sigma \}$ and the MC2 levels by 
$\{J,w\}$.  In this section we focus on the numerous levels that lie beneath the 
$j=2$ asymptote as they are most sensitive to the short-range potentials.

\subsection{Single channel}

\begin{figure}
\begin{center}
\includegraphics[width=0.8\textwidth]{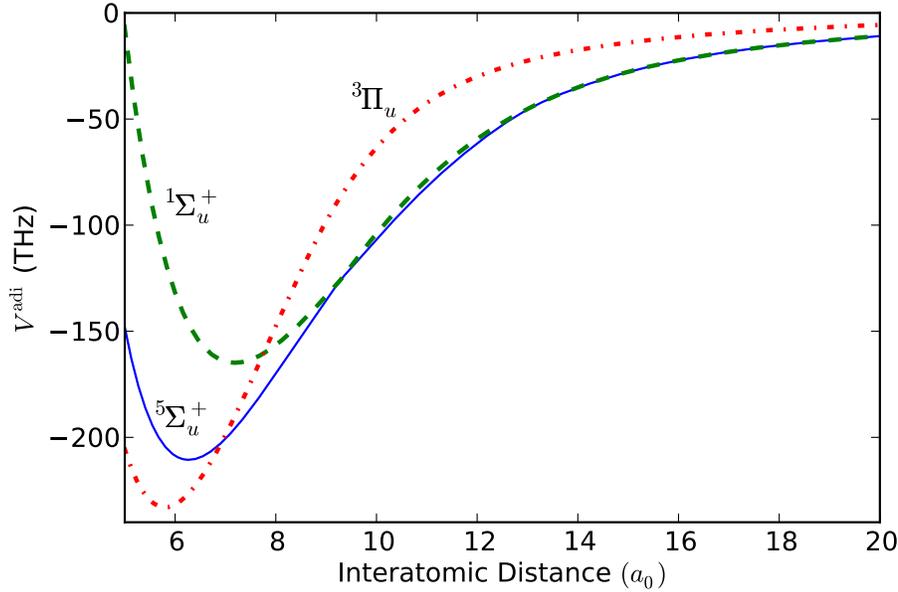}
\caption{\label{scpots}Single-channel potentials of the $0_u^+$, $J=1$ manifold
and their short-range case (a) assignments. The near degeneracies in the $^{1,5}\Sigma^+_u$ 
potentials over the region $17\,a_{0} < R < 17.5\,a_{0}$ affect the determination of
potential crossings in this region.}
\end{center}
\end{figure}

Our single-channel results for levels associated with the 2s$\,{}^3$S$_1$ + 
2p$\,{}^3$P$_2$ asymptotic limit are presented in table \ref{tab:adi} and 
compared to those of \cite{DLGD09}.  Of the eight tabulated series in 
\cite{DLGD09}, six, including the $1_u$ and $2_u$ series presented in table 
\ref{tab:adi}, match very well.  Almost all results are within 0.3\% of the 
values tabulated in \cite{DLGD09} and the maximum absolute difference is $20$ 
MHz which is comparable to experimental accuracy.  The two series that disagree 
are $0_u^+$, $J=1,3$.  The differences arise from ambiguities in smoothly 
connecting the diagonalized potentials where the two nearly degenerate BO 
potentials ${}^{1,5}\Sigma^+_u$ cross in the region $17\,a_{0} < R < 
17.5\,a_{0}$ (see figure \ref{scpots}).  Applying a fine diagonalization grid for these sets leads to 
avoided crossings in the single-channel potentials whereas a coarser grid gives 
rise to crossings.  The values presented in table \ref{tab:adi} were obtained 
using a fine grid of $\Delta R_{n}=0.01\,a_0$.  We note that $1_{g}$ was the 
only other single-channel set that exhibited this dependence upon the 
diagonalization grid, due to crossings in the nearly degenerate ${}^{1,5}\Pi_g$ 
BO potentials.

\begin{table}
\caption{\label{tab:adi} Binding energies of the single-channel levels in units 
of GHz. All potentials are asymptotic to $j=2$, were diagonalized using a grid 
spacing of $0.01\,a_0$, and have minima of approximately $-211$ THz.}
\begin{indented}
\lineup
\item[]\begin{tabular}{lllllll}
\br
 & \centre{2}{$1_u$, $J=1$} & \centre{2}{$2_u$, $J=2$} & \centre{2}{$0_u^+$, $J=1$} \\
\ns\ns & \crule{2} & \crule{2} & \crule{2} \\
$v$ & This work & Ref \cite{DLGD09} & This work & Ref \cite{DLGD09} & This work & Ref \cite{DLGD09} \\
\mr
70 & 11.319  & 11.301   & 13.666    & 13.647   &  14.583   & 13.658  \\
71 & \07.167 &  \07.154 &  \09.020  &  \09.006 &  \09.699  &  \09.029  \\
72 & \04.316 &  \04.307 &  \05.703  &  \05.692 &  \06.205  &  \05.735  \\
73 &  \02.432 &  \02.426 &  \03.414 &  \03.407 &  \03.796  &  \03.481  \\
74 &  \01.250 &  \01.246 &  \01.913 &  \01.908 &  \02.215  &  \02.015  \\
75 &  \00.566 &  \00.564 &  \00.996 &  \00.992 &  \01.229  &  \01.108  \\
76 &  \00.216 &  \00.215 &  \00.478 &  \00.476 &  \00.642  &  \00.572  \\
\br
\end{tabular}
\end{indented}
\end{table}

\subsection{Multichannel without Coriolis}

Next we compare the single-channel results with those from the multichannel 
calculations that do not include Coriolis couplings.  These methods differ only 
by the introduction of the non-adiabatic couplings.  Although a multichannel 
calculation introduces new complexities into the analysis, additional benefits 
arise from the ability to analyze the multichannel wave functions belonging to 
each level.

For the majority of levels there are negligible differences between the results 
from the two methods.  The results for the $0_u^-$, $0_g^\pm$, $1_u$, $2_g$, 
$2_u$ and $3_g$ case (c) sets agree to better than $0.1\%$ for all but the least 
bound levels where the differences are less than $1\%$. These differences are 
smaller than the uncertainties in the potentials and are much smaller than the 
uncertainties in experimental results. There are no bound levels for the $3_u$ 
set.

By contrast the $0_u^+$ and $1_g$ case (c) sets include results that differ 
significantly from the single-channel results.  
These differences were isolated to two of the four 
single-channel potentials from the $0_u^+$ set and three of the six 
single-channel potentials from the $1_g$ set. For the $0_u^+$ sets, the 
differences were up to $2\%$ for binding energies greater than $2$ GHz and up to 
$5\%$ otherwise. For the $1_g$ sets the results differed by up to $15\%$.  The 
levels for the remaining potentials differ by less than $0.1\%$.  We surmise 
that the near degeneracies in the interacting adiabatic potentials are the cause 
for these differences.  MC1 and SC results for the case (c) sets $2_u$, $J=2$ 
and $0_u^+$, $J=1$ are presented in table \ref{tab:MC1}. 

\begin{table}
\caption{\label{tab:MC1} Binding energies of the $0_u^+$, $J=1$ and $2_u$, $J=2$ 
levels in the potential asymptotic to $j=2$ in units of GHz obtained from 
multichannel calculations without Coriolis terms (MC1) and from single-channel 
calculations.  The $2_u$ single-channel potentials labelled A and B have minima 
of approximately $-211$ THz and $-233$ THz respectively and the $0_u^+$ 
single-channel potentials labelled A and B have minima of approximately 
$-211$~THz and $-165$~THz.}
\begin{indented}
\item[]\begin{tabular}{lllllllllll}
\br
\centre{5}{$2_u$, $J=2$} & \centre{6}{$0_u^+$, $J=1$} \\  \bs
\crule{5} & \crule{6} \\  \bs
 \centre{2}{MC1} & \centre{2}{Single-Channel} &  & \centre{2}{MC1} &
\centre{3}{Single-Channel} & \\
$v$ &  & $j=2$ & $j=2$ & Diff. & $v$ &  &
$j=2$ & $j=2$ & $j=1$ & Diff. \\
& & A & B & (\%) & & & A & B & & (\%) \\
\mr
125 & 20.010 & 20.008 & & 0.0 & 189 & 19.511 & & & 19.556 & 0.2 \\
126 & 19.906 & & 19.907 & 0.0 & 190 & 16.115 & & 16.085 & & 0.2 \\
127 & 13.667 & 13.666 & & 0.0 & 191 & 14.565 & 14.583 & & & 0.1 \\
128 & 12.346 & & 12.348 & 0.0 & 192 & 10.543 & & 10.491 & & 0.5 \\
129 & 9.0227 & 9.0230 & & 0.0 & 193 & 9.6694 & 9.6986 & & & 0.3 \\
130 & 7.2495 & & 7.2535 & 0.1 & 194 & 8.7252 & & & 8.7776 & 0.6 \\
131 & 5.7068 & 5.7027 & & 0.1 & 195 & 6.6048 & & 6.5794 & & 0.4 \\
132 & 3.9174 & & 3.9253 & 0.2 & 196 & 6.1999 & 6.2046 & & & 0.1 \\
133 & 3.4205 & 3.4141 & & 0.2 & 197 & 3.8897 & & 3.9063 & & 0.4 \\
134 & 1.9247 & 1.9132 & & 0.6 & 198 & 3.8223 & 3.7960 & & & 0.7 \\
135 & 1.8065 & & 1.8231 & 0.9 & 199 & 2.2539 & 2.2154 & & & 1.7 \\
136 & 1.0050 & 0.9956 & & 0.9 & 200 & 2.0855 & & 2.1245 & & 1.9 \\
137 & 0.5395 & & 0.5566 & 3.1 & 201 & 1.8130 & & & 1.8620 & 2.7 \\
138 & 0.4824 & 0.4776 & & 1.0 & 202 & 1.2618 & 1.2292 & & & 2.6 \\
139 & 0.2113 & 0.2076 & & 1.8 & 203 & 0.9663 & & 1.0009 & & 3.5 \\
\br
\end{tabular}
\end{indented}
\end{table}

The relative proportions of each single-channel basis in the multichannel bound 
state eigenfunction can be calculated by applying the unitary transformation 
$D_{ak}(R)$ defined in equation (\ref{cwp30b}).  For the majority of levels only 
one single-channel basis state is present, indicating that there is minimal 
difference between the single- and multi- channel methods.  Levels in the 
$0_u^+$ and $1_g$ sets, however, have multiple single-channel contributions, 
reinforcing the observation that their binding energies differ markedly between 
the methods.

\subsection{Multichannel with Coriolis}

The final comparison we make is between the two multichannel methods, where the 
only difference is the inclusion or otherwise of the Coriolis couplings.  
Although the Coriolis couplings vary as $R^{-2}$ and might be expected to play a 
significant role at small $R$, we do not see any difference for the more deeply 
bound levels.  The majority of levels for the $J=1,2$ ungerade sets differ 
between the MC1 and MC2 results by less than $0.1\%$ and for detunings larger 
than $7$~GHz from the $j=2$ asymptote, all results agree to within $0.5\%$.  
These differences are slightly greater than those between the single-channel and 
multichannel results without Coriolis couplings. The $J=3$ set however shows 
much more variance.  For detunings from the $j=2$ asymptote of more than 
$5000$~GHz, the results differ by $< 0.5\%$ but some of the more shallowly bound 
levels, with detunings larger than $15$~GHz, differ by up to $2\%$.  Levels with 
$J>3$ were not investigated as selection rules forbid excitation to these levels 
during $s$-wave collisions in the metastable state.

The most weakly bound levels with wavefunctions that extend from the small to 
large interatomic regions show the greatest effects of including Coriolis 
couplings.  For the $J=1,2$ sets, approximately a third of the levels for 
detunings less than $7$~GHz show differences ranging from $1\%$ to $5\%$, which 
are larger than the experimental uncertainties. The $J=3$ set shows even larger 
differences, with levels that can differ by up to $10\%$ for detunings smaller 
than $15$~GHz. These differences reflect the increase of Coriolis couplings with 
$J$ and are quite significant, especially as the available experimental 
measurements are almost entirely within this range.

The gerade $J=1,2,3$ sets have similar behaviour to the ungerade sets and show 
deviations up to $0.5\%$ between the methods for detunings larger than $800$, 
$3600$ and $5000$ GHz respectively.  For smaller detunings, differences up to 
$10\%$ do occur, although the differences for most levels remain within $0.5\%$.  
A sample comparison of the results for gerade levels obtained using multichannel 
calculations with and without Coriolis couplings is presented in table 
\ref{tab:MC2}.

The differences in binding energy can be separated into two components; a shift 
that occurs for all levels and a splitting that occurs for degenerate MC1 
$\Omega_{j} = \pm \Omega$ levels. The splittings between these levels in the MC2 
results are generally less than $0.1\%$ for the $J=2$ ungerade set although 
several are close to $1\%$. For the remaining sets, the Coriolis splittings 
steadily increase as the detuning decreases, with most less than $2\%$. The 
greatest splittings occur when MC1 levels of different $\Omega$ manifolds are 
closely spaced and can be up to $10\%$.  The only set that does not contain 
either the $0_u^+$ or $1_g$ basis is the ungerade $J=2$ set which shows the 
least splitting of the degenerate MC1 levels.  Therefore we conclude that near 
degeneracies in the adiabatic potentials may be the cause for strong splitting 
as was observed for the differences between the SC and MC1 levels.

\begin{table}
\caption{\label{tab:MC2} Binding energies of the gerade $J=2$ set in the 
potential asymptotic to $j=2$ in units of GHz obtained from multichannel 
calculations with (MC2) and without (MC1) Coriolis couplings.  Note that levels 
with $\Omega > 0$ are doubly degenerate in the MC1 coupling scheme.}
\begin{indented}
\lineup
\item[]\begin{tabular}{clcccc}
\br
 \centre{2}{MC2} & \centre{3}{MC1} & \\ \bs
\crule{2} & \crule{3} & \\
$v$ &  & $0_g^+$ & $1_g$ & $2_g$ & Diff.(\%) \\
\mr

601 & 4.1293 & & 4.1474 & & 0.4 \\
602 & 4.1287 & & 4.1474 & & 0.5 \\
603 & 3.9403 & & & 3.9174 & 0.6 \\
604 & 3.9400 & & & 3.9174 & 0.6 \\
605 & 3.4378 & & & 3.4205 & 0.5 \\
606 & 3.4378 & & & 3.4205 & 0.5 \\
607 & 3.3656 & & 3.3889 & & 0.7 \\
608 & 3.3536 & & 3.3889 & & 1.0 \\
609 & 2.6660 & 2.6552 & & & 0.4 \\
610 & 2.2982 & & 2.3116 & & 0.6 \\
611 & 2.2976 & & 2.3116 & & 0.6 \\
\br
\end{tabular}
\end{indented}
\end{table}

The contribution from each case (c) subspace for each level in the full 
multichannel calculation can be determined from the bound state eigenfunction.  
We find no obvious correlation between these case (c) proportions and the 
differences in the binding energies, but do observe that most levels possess 
large contributions from two or more case (c) sets, despite the relatively small 
differences in binding energies for the deeply bound states.  The ungerade 
$J=1,2$ sets do possess some weakly bound levels that occupy the largest 
interatomic distances and these are dominated by contributions of greater than 95\% 
from one case (c) set only.

\subsection{$j=1$ and $j=0$ asymptotes} \label{sec:resonance}

Calculations of energy levels near the dissociative limits 2s$\,{}^3$S$_1$ + 
2p$\,{}^3$P$_j$  for $j=1$ and $j=0$ introduce open channels corresponding to 
$j=2$ and $j=1,2$ respectively into the multichannel calculation. The 
renormalized Numerov method is again used to find the allowed energies of the 
system. Due to the open channels, true bound levels appear amongst a set of 
artificial box states that represent the continuum of scattering states after an 
outer boundary wall is imposed.  These artificial states are identified by 
examination of the wave function, or by their variability as the outer boundary 
is repositioned.

We find very few bound levels with purely real binding energies; too few to 
explain the experimental measurements. Hence, we search for any resonances that 
predissociate via the open channels.  The search to find the complex resonance 
energies that smoothly match the solutions obtained by inward and outward 
integration \cite{Johnson78} is performed by making use of Cauchy's argument 
principle, which states that a meromorphic function $f(z)$ can be shown to 
satisfy
\begin{equation}
\label{cwp31a}
\frac{1}{2\pi i} \oint_C \frac{f^\prime(z)}{f(z)} \rmd z = N - P \,,
\end{equation}
where $C$ is a bounded contour which contains $N$ zeros and $P$ poles.  
We choose the function to be the determinant 
$D(E)= |\mathbf{R}_{m}-\hat{\mathbf{R}}^{-1}_{m+1}|$ where $\mathbf{R}_{m}$ and
$\hat{\mathbf{R}}_{m+1}$ are ratio matrices for the outward and 
inward integrations respectively of the renormalized Numerov method, 
see equations (21) and (25) of \cite{Johnson78}. The matching condition
is then $D(E)=0$. If $D(E)$ is assumed to be 
meromorphic then a region bounded by 
a contour should show $N$ zeros for $N$ valid resonances.  Because poles in the 
same region will `cancel' zeros then it is important to use relatively small 
contours such that each region can be assumed to contain only a single zero or 
pole.  Additionally, the matching point $R_m$ for the function $D(E)$  must be 
placed so that it is within the classically permitted $R$ region of the resonance 
state wave function.  For resonances that extend into short interatomic distances a 
single value of $R_m = 100\;a_{0}$ is sufficient to lie within the outer turning point
of these resonances. Long-range resonances, however,  
require multiple scans at different $R_m$ values in order to detect all 
resonances because the classically allowed regions of the different wave functions may not overlap.  
The results in this paper originated from scans using matching 
distances of $100,\,200,\,400$ and $600\;a_0$.

The numerical procedure uses the argument principle with a box contour that has 
a real width of $5$ MHz, a upper imaginary boundary of $1$ MHz and a lower 
imaginary boundary of $-100$ MHz. This places a limit upon the maximum 
predissociation that will be detected as a valid resonance. A range of real 
energies is scanned by using many adjacent box contours.  The derivative 
$D^\prime(E)$ in each integral is easily calculated numerically and the integral 
itself is evaluated using an adaptive Gauss-Kronrod quadrature method.  This 
allows for a high density of grid points when required in parts of the integral 
and can hence handle the presence of poles and zeros in the neighbourhood of the 
contours.

Once regions are found that are known to contain valid resonances, the box 
contour of each region is narrowed using a bisection method extended to two 
dimensions. Because the integrals in (\ref{cwp31a}) are computationally 
expensive to calculate, a gradient descent method is used to determine the exact 
resonance energy once the box edges are smaller than 1 MHz.

The single-channel results for $0_g^-$ near the $j=1$ asymptote agree well with 
those tabulated in \cite{DLGD09} to within 4 MHz. The full multichannel 
calculation for resonances that extend into the short-range region, that is, all 
resonances that are not purely long-range, modifies the resonance energies near 
the $j=1$ asymptote by up to $5$~MHz for the $J=1$ sets, increasing to $20$~MHz 
for the $J=3$ sets, which is similar to the behaviour observed near the $j=2$ 
asymptote.  No short-range resonances are found near the $j=0$ asymptote.

The multichannel results for the purely long-range $1_{g}$ and $0^{+}_{u}$ 
levels near the $j=0$ dissociation limit and the purely long-range $0^{-}_{u}$ 
and $2_{u}$ levels  near the $j=1$ dissociation limit differ by less than $1$ 
MHz from the multichannel results of \cite{Venturi03}.  This is to be expected 
as these levels do not depend on the new short-range potentials. We have also 
examined the $1_{g},J=1$ levels in the $j=1$ asymptote that \cite{Venturi03} 
rejected as purely long-range on the grounds that barrier penetration through 
the double well structure of the potentials leads to significant probability for 
$R < 200\,a_{0}$ and that there appeared to be appreciable predissociation. We 
find three pairs of resonances, $(-165.0,-164.4)$, $(-63.0,-62.7)$ and
$(-16.2,-15.9)$ MHz, which do indeed possess significant short-range probability,
however their predissociation widths are relatively small, ranging from 0.23
to 1.08 MHz. Each pair represents the splitting of the $\Omega_{j} = \pm \Omega$
degeneracy by the Coriolis couplings.

\section{Experimental Assignment}

\subsection{Basis of assignments}

Previous attempts to assign quantum numbers to experimental measurements have 
primarily been motivated by the desire to minimize the differences between the 
theoretical and experimental binding energies. In addition, knowledge of the 
short-range character of the individual levels is useful as this determines the 
likelihood of Penning ionization.  Acting as both a level broadening mechanism 
and a detection method for most experiments, Penning ionization is extremely 
likely to occur if the bound level possesses singlet or triplet character in its 
Hund's case (a) basis for $R < 20$~$a_0$.  We follow \cite{Leonard05} in 
assuming that this probability is unity for a level with only singlet or triplet 
short-range character.  Although  single-channel calculations cannot determine 
accurately the short-range character of each bound level, L\'eonard \etal 
\cite{Leonard05} have used the absence or presence of different experimental 
levels to infer the rotational coupling between different case (c) subsets.

The multichannel calculations in this paper enable the complete bound state 
eigenfunctions to be calculated and the short-range spin-$S$ fraction 
$f_{2S+1,v}$ to be determined exactly for each level, see 
\ref{app:decomposition}.  To assist in assigning the experimental levels we can 
also calculate an approximate coupling factor $\mathcal{A}_{v}$ between the 
excited bound state $|\Psi \rangle = \sum_a R^{-1}G_{a,v}(R)|a \rangle$ and the 
initial metastable ground states $R^{-1}G_g(R)|Sm_{S}lm_{l}\rangle $ where we 
suppress the labels $L_{1}=m_{L_{1}}=L_{2}=m_{L_{2}}=0$. This factor is 
estimated from the Franck-Condon integral
\begin{equation}
\label{cwp32}
\mathcal{A}_{v}= \frac{1}{N_g} \sum_{a,g} \langle S m_S l m_l| 
\hat{H}_\mathrm{int} | a\rangle \int G_{g}(R)G_{a,v}(R) \rmd R \,,
\end{equation}
where the matrix element of the laser interaction $\hat{H}_\mathrm{int}$ is 
given in \ref{app:coupling} and $N_g$ represents the number of coupled 
metastable states.  We replace the radial eigenfunction $G_{g}(R)$ for the 
metastable state by unity, as we do not wish to specify the temperature of the 
system which can range from $\mu$K to mK.  The $\Omega$-degenerate levels that 
have been split by the Coriolis coupling form pairs, each comprised of a 
symmetric and an antisymmetric level.  As only the symmetric level of each pair 
can possess a non-zero coupling, only one of the pair of Coriolis split levels 
can possibly be observed by experiment.

\subsection{Spin-polarized experiments for $j=2$ \label{sec:polarized}}

We first consider the experimental data of Kim \etal \cite{Kim04} for 
spin-polarized metastable atoms in a magnetic trap at approximately 10~$\mu$K. 
The experiment used an optical detection method and was able to measure 
separately the number of atoms, the optical density and the temperature of the gas 
after a cycle of laser pulse, thermalization and ballistic expansion. As the 
method is based upon thermalization, any decay path from the excited state is 
detected, including spontaneous emission.  In the experiment, systematic scans 
for detunings in the range $0-6$~GHz from the $j=2$ asymptote were performed and 
additional narrow scans centred upon predictions for more deeply bound levels 
were carried out.  Only ungerade excited levels are accessible from the 
spin-polarized metastable state $|22lm_{l}\rangle$ and higher partial waves than 
$s$-waves will contribute little to collisions at 10~$\mu$K.

The theoretical calculations find a possible 118 levels with binding energies in 
the range $0.08-6$~GHz. To identify those levels that should result in strong, 
narrow and experimentally detectable resonances, we impose conditions on the 
maximum probability of Penning ionization by insisting that the quintet
short-range character be at least 87.5\% , and on the minimum coupling to the 
metastable manifold by requiring $\mathcal{A}_{v} \geq 0.9\;E_{h}$, see 
(\ref{cwp32}).  These conditions 
isolate 19 levels, shown in table \ref{tab:kim}, that match almost uniquely with 
experimental observations. In addition, another four levels may be assigned to 
observations by relaxing the constraint upon one of the observability criteria;
this may be justified by considering the strength of the other observability 
criterion. Only one experimental observation of $0.455$ GHz remains unassigned.  
Two assignments of very close binding energies have been made to the observed 
level near $0.200$ GHz, and the closeness of these levels may mean they are 
impossible to distinguish as separate in experiment. Also shown in table 
\ref{tab:kim} are the case (c) contributions of greater than 20\% to each level, 
listed in descending order of contribution.

Experimental levels above 6 GHz have not been included in the initial 
assignments as we would have been unable to properly determine the observability
criteria without an unbroken scan region.  However, the theoretical levels that 
are predicted to be observable in the range $6-14$ GHz have been appended to 
table \ref{tab:kim}, along with assignments to experimental observations if 
appropriate.

\renewcommand{\dash}{\multicolumn{1}{c}{---}}
\begin{table}
\caption{\label{tab:kim}Theoretical levels that are predicted to be 
experimentally observable and their assignment to the experimental data of 
\cite{Kim04}.  Energies are given in GHz relative to the $j=2$ asymptote. The 
third column lists the levels after a 1\% variation is applied to the 
short-range potentials and the fourth and fifth columns are the observability 
quantities calculated from the results of this variation.}
\begin{indented}
\lineup
\item[]\begin{tabular}{lllllll}
\br
Exp & Theor & Variation & $\mathcal{A}_{v}\;(E_{h})$ & $f_{5,v}(\%)$ &  Case (c) & \cite{DLGD09} assignment \\
\mr
\-5.90  & \-5.729 & \-5.920 & 1.298 & 99.9 & $2_u,\;J=2$ & $2_u,\; J=2$ \\
\-5.64  & \-5.463 & \-5.648 & 1.770 & 97.5 & $2_u, \;J=3$ & $0_u^+,\; J=1$ \\
\-4.53  & \-4.394 & \-4.551 & 1.129 & 97.7 & $1_u,0_u^+,\; J=1$ & $0_u^+, \;J=3$ \\
\-4.25  & \-4.142 & \-4.285 & 1.876 & 93.1 & $0_u^+,1_u, \;J=3 $ & $1_u,\; J=2$ \\
\-3.57  & \-3.438 & \-3.566 & 1.394 & 99.3 & $2_u, \;J=2 $ & $0_u^+,\; J=1 $\\
\-3.37  & \-3.251 & \-3.375 & 2.023 & 98.2 & $2_u,\; J=3 $ & $2_u,\; J=2$ \\
\-2.59  & \-2.499 & \-2.603 & 1.143 & 99.1 & $1_u,0_u^+,\; J=1 $ & $1_u,\; J=3$ \\
\-2.42  & \-2.338 & \-2.433 & 2.167 & 96.6 & $0_u^+,1_u, \;J=3$ & $1_u,\; J=1$ \\
\-2.00  & \-1.937 & \-2.019 & 1.431 & 99.4 & $2_u,\; J=2$ & $0_u^+,\; J=1 $\\
\-1.88  & \-1.807 & \-1.886 & 2.399 & 99.8 & $2_u,\; J=3$ & $2_u,\; J=2$ \\
\-1.37  & \-1.326 & \-1.387 & 2.399 & 99.3 & $0_u^+,1_u,\; J=1$ & $1_u,\; J=3$ \\
\-1.275 & \-1.223 & \-1.282 & 2.504 & 99.7 & $0_u^+,1_u,\; J=3$ & \dash \\
\-1.22  & \-1.160$^{\rm a}$  & \-1.220$^{\rm a}$ & 0.565 & 100 &  $1_u,\; J=2$ & $1_u, \;J=2$ \\
\-1.07  & \-1.013 & \-1.062 & 1.546 & 99.6 & $2_u,\; J=2$ & $0_u^+,\; J=1$ \\
\-0.98  & \-0.928$^{\rm a}$ & \-0.973 & 2.957 & 99.2 & $2_u,\; J=3$ & $2_u,\; J=2$ \\
\-0.62  & \-0.589 & \-0.621 & 4.435 & 92.5 & $0_u^+,2_u,\; J=3$ & $1_u,\; J=1 $\\
\-0.51  & \-0.487 & \-0.511$^{\rm a}$ & 1.459 & 63.5 & $2_u,\; J=2$ & $2_u,\; J=2$ \\
\-0.455 & \dash   & \-0.458 & 4.391 & 90.5 & $2_u,0_u^+, \; J=3$ & $0_u^+, J=3$ \\
\-0.280 & \-0.263$^{\rm a}$  & \-0.276$^{\rm a}$ & 5.086 & 82.3 & $0_u^+,2_u,\; J=3$ & $0_u^+, J=1$ \\
\-0.235 & \-0.223$^{\rm a}$  & \-0.228 & 1.813 & 99.3 &  $2_u,\; J=2$ & $1_u,\; J=1$ \\
\-0.200 & \-0.185 & \-0.199 & 0.993 & 99.9 & $1_u,\; J=2$ & $2_u,\; J=2 $\\
\dash   & \-0.184 & \-0.196 & 3.283 & 93.2 & $2_u,\; J=3$ & \dash \\
\-0.185 & \-0.167 & \-0.178$^{\rm a}$ & 4.691 & 86.1 & $0_u^+,1_u,\; J=3$ & $1_u,\; J=2$ \\
\-0.09  & \-0.083 & \-0.089 & 2.287 & 99.8 & $2_u,\; J=2$ & $0_u^+,\; J=3$ \\
\\
\hline
\\
\-13.67        & \-13.259 & \-13.621 & 1.376 & 97.5 & $2_u$, $J=3$ & $2_u$, $J=2$ \\
\-11.70        & \-11.434 & \-11.764 & 1.057 & 96.8 & $1_u$, $J=1$ & $1_u$, $J=1$ \\
\multicolumn{1}{c}{---$^{\rm b}$} & \-9.051  & \-9.324  & 1.201 & 100  & $2_u$, $J=2$ & \dash \\
\-8.95         & \-8.705  & \-8.967  & 1.557 & 97.6 & $2_u$, $J=3$ & $2_u$, $J=2$ \\
\-7.45         & \-7.262  & \-7.493  & 1.096 & 97.0 & $1_u$, $J=1$ & $1_u$, $J=1$ \\
\br
\end{tabular}
\item[]$^{\rm a}$Observability criteria relaxed.
\item[]$^{\rm b}$Region not scanned in experiment.
\end{indented}
\end{table}

The theoretical levels that match have consistently lower binding energies than 
their assigned experimental levels.  This suggests that a small correction to 
the input short-range potentials is required.  Motivated by the observation in 
\cite{Leo01} that many ultracold molecular properties are very sensitive to the 
slope of the potential at the inner classical turning point and the fact that 
the observable levels possess strong quintet short-range character, we choose to 
vary the quintet MRCI potentials in this region by introducing a multiplicative 
factor $c$ through the smoothing function
\begin{equation}
\label{cwp32a}
V^{\prime}(R) = \cases{ V(R) (1+2c) & $R < R_1$ \\
				V(R) \left[ 1 + c(1 + \cos a (R - R_1)) \right] & $R_1 < R < R_2$ \\
				V(R) & $R_2 < R$}\,,
\end{equation}
where $R_1=5\,a_0$, $R_2=10\;a_0$ and $a = \pi / (R_2-R_1)$.  The value 
$c=0.005$ that represents a 1\% variation which is quickly turned on through the 
region $5 < R < 10\,a_{0}$, deepens the minima of the attractive ${}^5 \Pi_u$ 
potential at $R = 5.387\;a_0$ by $0.985\%$ and moves it to a closer interatomic 
distance by $0.003\;a_0$. The only other ungerade quintet potential ${}^5 
\Sigma^+_u$ is not significantly affected as it is repulsive.  These varied 
potentials produced much better agreement with experiment and most levels are 
within the experimental uncertainty of 20~MHz, see table \ref{tab:kim}. A 
further improvement of the observability parameters for levels $0.928$ and 
$0.228$ GHz means that a relaxation of the criteria is no longer required in 
those cases and the observation of $0.455$ GHz can be assigned to a theoretical 
level of $0.458$ GHz.  Unfortunately, the levels at $0.511$ and $0.178$ GHz 
require a relaxation of the criteria under the short-range variation.

The reason for the very close match between the theoretical and experimental 
levels resulting from the short-range variation is unclear.  Variation of the 
non-quintet MCSCF potentials only, which are more likely to possess small 
inaccuracies than the MRCI quintet potentials \cite{Dickinson09}, barely affects 
the binding energies and in some instances adversely affects the observability 
parameters. Variation of all the potentials however, produces a similar 
improvement in the binding energies of the levels but does not allow for the 
assignment to the observed level at $0.455$ GHz.

There is a large difference between the assignment of levels obtained from our 
procedure and from that of \cite{DLGD09}. The assignment procedure of 
\cite{DLGD09} rejected entire adiabatic series by characterizing the likelihood 
of Penning ionization of the adiabatic potentials, whereas our multichannel 
calculation allows us to reject levels on an individual basis.  Our calculations 
indicate that very few levels of each series are accessible by photoassociation, 
implying that the logical inclination to attempt to complete an entire adiabatic 
series when matching to experimental measurements appears to be invalid.  
Interestingly, our assignments agree with those of \cite{Leonard05} for all 
levels except those for which we assign two case (c) sets and those with 
energies of $-2.42$ GHz and $-1.22$ GHz.

\subsection{Unpolarized experiments for $j=2$ \label{sec:unpolarized}}
 
Next, we attempt to assign levels to the experimental results of van Rijnbach 
\cite{Rijn04} for unpolarized metastable atoms in a magneto-optical trap 
at temperatures of approximately 2 mK, where detection of Penning ionization was 
used as the spectroscopic signal. The temperature and lack of polarization 
allows a large number of metastable collisional channels to contribute, and both 
gerade and ungerade excited states are accessible, although the dipole 
approximation requires that the gerade excited states can only be coupled to 
ungerade $l=1$ metastable levels.  The $l=1$ centrifugal barrier is significant 
for these ultracold experiments as the region $R < 240$~$a_0$ becomes 
classically forbidden for colliding metastable atoms at a temperature of 1~mK.

We find a total of 345 levels with energies between $-13.57$ GHz and $-0.045$ 
GHz.  A subset of levels is again isolated by imposing a condition upon the 
minimum required coupling to the metastable manifold given by $\mathcal{A}_{v} > 
0.7$~$E_h$.  In order to have a non-negligible Penning ionization signal for 
detection but to still impose an upper limit on the observable Penning 
ionization decay rate, we limit the quintet short-range character to $0.875 < 
f_{5,v} < 0.998$.  The $22$ levels that satisfy these conditions are shown in 
table \ref{tab:rijnbach} with their corresponding assignments.

There is reasonable agreement between the theoretical and experimental results, 
and again consistently smaller theoretical binding energies are observed.  
Applying the same short-range potential correction as in the previous section 
yields much improved agreement with the experimental observations and permits 
the inclusion of four additional assignments. One level at $-0.167$ GHz that was 
previously unassignable is now excluded.  Few levels, however, are within the 
tight experimental uncertainties of 0.002 to 0.19 GHz.  The experimental data of 
van Rijnbach includes 15 levels that were very weakly observed and 11 of these 
have not been given assignments in table \ref{tab:rijnbach}.  To do so requires 
a relaxation of our imposed theoretical conditions which would introduce many 
additional theoretical results that cannot be assigned to any experimental 
level. The only `strong'  experimental level unassigned is $-2.87$ GHz, although 
a relaxation of the criteria is required to make an assignment to $-0.52$ and 
$-0.27$ GHz.  The theoretical levels of $-1.062$ and $-0.228$ GHz satisfy our 
imposed conditions but do not correspond to any experimental level of 
\cite{Rijn04}.   We also note that no gerade levels are present in the 
assignments, although one level at a detuning of $-0.109$ GHz does appear in the 
initial calculations using the unvaried potentials.  The absence of gerade 
levels is due not to the $l=1$ centrifugal barrier in the metastable state as we 
do not specify the wave function for the metastable states in the approximate 
coupling factor $\mathcal{A}_v$ but rather to the spin-conserving nature of the 
coupling. The $l=1$ metastable channels are in the triplet configuration and 
will strongly couple to excited levels with large triplet character, levels that 
are very likely to undergo Penning ionization.  Almost all of the 39 gerade 
levels with binding energies in the range 0.045 to 13.57 GHz which satisfy 
$\mathcal{A}_v > 0.7\,E_{h}$ have $f_{5,v} < 0.2$. This, together with the large 
$l=1$ centrifugal barrier, means that the entire gerade manifold cannot be 
considered for assignment to the experimental levels near the $j=2$ asymptote.

The measurements of Tol \cite{Tol05} for unpolarized atoms at a temperature of 
$1$ mK overlap for nearly all levels with the measurements of \cite{Rijn04} 
listed in table \ref{tab:rijnbach}.  
However, there are two additional levels, one at $-0.622$ GHz which fills a gap 
in the table corresponding to the theoretical level $-0.621$ GHz and the other 
at $-0.045$ GHz which can be assigned to a theoretical level with energy 
$-0.055$ GHz.

If the short-range variation is applied to all of the potentials, instead of 
just to the quintet MRCI potentials, similar agreement is obtained in the shift 
of binding energies. More importantly, however, the behaviour of the 
observability parameters does not change significantly, in contrast to the 
beneficial changes shown in table \ref{tab:rijnbach} arising from varying only 
the quintet potentials.

There is again a large difference between the assignment of levels using our 
procedure as opposed to that of \cite{DLGD09}.  L\'eonard \etal \cite{Leonard05} 
have argued that the $1_{u}$ and $2_{u}$ $J=2$ and all the $0^{+}_{u}$ 
assignments should not appear as they are not expected to produce ions in the 
adiabatic approximation.  However they have made two assignments to the $2_u$ 
$J=2$ set in the expectation that nearby overlapping levels may be the cause of 
these observations \cite{Leonard09}.  We do not find any nearby overlapping 
levels that are accessible according to our observability criteria and instead 
believe that the assignments must be made to the $2_{u}$ $J=2$ set, as the 
multichannel calculation firmly places these levels in the ion detection regime.  
We conclude that the Coriolis couplings introduce a marked change in the 
short-range character of these states that allows ion production to occur.  We 
find only one other level at $-3.49$ GHz that follows this behaviour and is 
assigned to $2_{u}$ $J=2$. Our remaining assignments agree with those of 
\cite{Leonard05} for all levels except those for which we assign two case (c) 
sets.

\begin{table}
\caption{\label{tab:rijnbach}Theoretical levels predicted to be experimentally 
observable and their assignment to the experimental data in \cite{Rijn04} and 
\cite{Tol05}.  Energies are given in GHz relative to the $j=2$ asymptote.  The 
third column lists the levels after a 1\% variation is applied to the 
short-range potentials and the fourth and fifth columns are the observability 
quantities calculated from the results of this variation.}
\begin{indented}
\lineup
\item[]\begin{tabular}{llllllll}
\br
Exp & Theor & Variation & $\mathcal{A}_v(E_h)$ & $f_{5,v}(\%)$ & Case (c) & \cite{DLGD09} assignment \\
\mr
\-13.57           & \-13.259 & \-13.621 & 0.978 & 97.5 & $2_u, \; J=3$ & $2_u, \; J=2$ \\
\-11.70$^{\rm b}$ & \-11.434 & \-11.764 & 0.700 & 96.8 & $1_u, \; J=1$ & $1_u, \; J=1$ \\
\-11.10$^{\rm b}$ & \-10.930 & \-11.224 & 1.042 & 89.5 & $1_u,0_u^+, \; J=3$ & $1_u, \; J=2$ \\
\-8.94            & \-8.705  & \-8.966  & 1.107 & 97.6 & $2_u, \; J=3$ & $2_u, \; J=2$ \\
\-7.44            & \-7.262  & \-7.493  & 0.733 & 97.0 & $1_u, \; J=1$ & $1_u, \; J=1$ \\
\-7.01            & \dash    & \-7.103  & 1.179 & 90.7 & $1_u,0_u^+, \; J=3$ & $1_u, \; J=2$ \\
\-5.64            & \-5.463  & \-5.647  & 1.258 & 97.5 & $2_u, \; J=3$ & $0_u^+, \; J=1$ \\
\-4.53            & \-4.394  & \-4.551  & 0.777 & 97.7 & $1_u,0_u^+, \; J=1$ & $0_u^+, \; J=3$ \\
\-4.26            & \-4.145  & \-4..285  & 1.333 & 93.1 & $0_u^+,1_u, \; J=3$ & $1_u, \; J=2$ \\
\-3.49$^{\rm b}$  & \-3.438  & \-3.566  & 1.214 & 99.3 & $2_u, \; J=2$ & $3_g, \; J=3$ \\
\-3.38            & \-3.251  & \-3.375  & 1.438 & 98.2 & $2_u, \; J=3$ & $2_u, \; J=2$ \\
\-2.87            & \dash    & \dash    & \dash & \dash & \dash & $2_u, \; J=4 $ \\
\-2.60            & \-2.499  & \-2.603  & 0.850 & 98.9 & $1_u,0_u^+, \; J=1$ & $1_u, \; J=3$ \\
\-2.42            & \-2.338  & \-2.433  & 1.541 & 96.6 & $0_u^+,1_u, \; J=3 $& $1_u, \; J=1$ \\
\-2.01            & \-1.937  & \-2.019  & 1.246 & 99.4 & $2_u, \; J=2 $& $0_u^+, \; J=1$ \\
\-1.88            & \-1.807  & \-1.886  & 1.705 & 99.8 & $2_u, \; J=3$ & $2_u, \; J=2$ \\
\-1.54            & \-1.326  & \-1.387  & 0.986 & 99.3 & $0_u^+,1_u, \; J=1$ & $2_u, \; J=4$ \\
\-1.28            & \-1.223  & \-1.282  & 1.780 & 99.7 & $0_u^+,1_u, \; J=3$ & $1_u, \; J=1$ \\
\dash             & \-1.013    & \-1.062  & 1.346 & 99.6 & $2_u, \; J=2$ & \dash \\
\-0.98            & \dash    & \-0.973  & 2.102 & 99.2 & $2_u, \; J=3$ & $2_u, \; J=2$ \\
\-0.622$^{\rm c}$ & \-0.589  & \-0.621  & 3.153 & 92.5 & $0_u^+,2_u, \; J=3 $& $1_u, \; J=1 $ \\
\-0.52            & \-0.487  & \-0.511$^{\rm a}$  & 1.270 & 63.5 & $2_u, \; J=2$ & $2_u, \; J=2 $\\
\-0.46            & \dash    & \-0.458  & 3.121 & 90.5 & $2_u,0_u^+, \; J=3$ & $0_u^+, \; J=3$ \\
\-0.27            & \dash    & \-0.276$^{\rm a}$ & 3.616 & 82.3 & $0_u^+,2_u, \; J=3$ & $0_u^+, \; J=1$ \\
\dash             & \dash    & \-0.228  & 1.578 & 99.3 & $2_u, \; J=2$ & \dash \\
\-0.19            & \-0.184  & \-0.196  & 2.334 & 93.2 & $2_u, \; J=3$ & $2_u, \; J=2$  \\
\dash             & \-0.167  & \dash    & \dash & \dash& $0_u^+,1_u, \; J=3$ & \dash \\
\-0.08$^{\rm b}$  & \-0.066  & \-0.071  & 2.968 & 99.6 & $2_u, \; J=3$ & $0_u^+, \; J=3$ \\
\-0.045$^{\rm c}$ & \-0.050  & \-0.055  & 4.056 & 98.1 & $0_u^+,1_u, \; J=3$ & $0_u^+, \; J=1$\\
\br
\end{tabular}
\item[]$^{\rm a}$Observability criteria relaxed.
\item[]$^{\rm b}$Weakly observed.
\item[]$^{\rm c}$Measurement of \cite{Tol05} that was not observed in \cite{Rijn04}.
\end{indented}
\end{table}

\subsection{Unpolarized experiments $(j=1)$}

Using the method described in section \ref{sec:resonance} to determine 
appropriate bound levels and resonances, we make assignments to the experimental 
observations of \cite{Rijn04} near the 
\mbox{He(2s$\,^{3}$S$_{1}$)+He(2p$\,^{3}$P$_{1}$)} asymptote. Although only the 
closed channels are used to discard artificial resonances, the properties of the 
resonances in table \ref{tab:j2} are calculated using the complete wavefunction.  
In addition to the observability criteria chosen previously to be $\mathcal{A}_v 
> > 0.7$~$E_h$ and $0.90 < f_{5,v} < 0.998$, a limit to the predissociation width 
$\Gamma_{\mathrm{pre}}$ of $100$~MHz is imposed as previously described in 
section \ref{sec:resonance}.  All resonances that satisfy these criteria are 
shown in table \ref{tab:j2} for detunings of less than 600 MHz from the $j=1$ 
asymptote.  Our assignments,  made by analysing contributions only from the 
closed channels that have asymptotes $j=1$ and $j=0$, agree well with those of 
\cite{DLGD09}.

\begin{table}
\caption{\label{tab:j2}Theoretical levels close to the $j=1$ asymptote and 
predissociation widths $\Gamma_{\mathrm{pre}}$ in MHz predicted to be 
experimentally observable and their assignment to the experimental data of 
\cite{Rijn04}.  Energies are given in GHz relative to the $j=1$ asymptote. The 
third column lists the levels after a 1\% variation is applied to the 
short-range potentials.  The parameters in the fourth, fifth and sixth columns 
are from the resonance calculated with variation.}
\begin{indented}
\lineup
\item[]\begin{tabular}{llllllll}
\br
Exp \cite{Rijn04} & Theor & Variation & $\Gamma_{\mathrm{pre}}$ & $\mathcal{A}_v(E_h)$ & $f_{5,v}(\%)$  & Case (c) & \cite{DLGD09} assignment  \\
\mr
\-0.452 & \-0.427 & \-0.466 & 75.4 & 0.770 & 98.7 & $0_g^-, \; J=3$ & $0_g^-, \; J=3$\\
\-0.343 & \-0.283 & \-0.309 & 7.7  & 1.800 & 98.6 & $0_g^-, \; J=1$ & $0_g^-, \; J=1$ \\
\-0.238 & \-0.182 & \-0.201 & 40.1 & 1.657 & 98.8 & $0_g^-, \; J=3$ & $0_g^-, \; J=3$ \\
\-0.159 & \-0.117 & \-0.128 & 2.3  & 2.082 & 91.1 & $0_g^-, \; J=1$ & $0_g^-, \; J=1$ \\
\-0.089 & \-0.065 & \-0.073 & 18.4 & 2.030 & 98.7 & $0_g^-, \; J=3$ & $0_g^-, \; J=3$ \\
\-0.043 & \-0.040 & \-0.045 & 1.4  & 2.943 & 99.5 & $0_g^-, \; J=1$ & $0_g^-, \; J=1$ \\
\br
\end{tabular}
\end{indented}
\end{table}

We note that, despite our rejection of the gerade manifold for assignments near 
the $j=2$ asymptote, all of these assignments near the $j=1$ asymptote are of 
the gerade manifold.  This may be explained by examining the adiabatic 
potentials and their triplet character over the entire interatomic range. 
If a gerade level is to be effectively coupled to the metastable manifold, 
it must possess triplet character at large 
interatomic distance yet become mostly quintet at short interatomic distance to 
avoid loss from ionization. The only two adiabatic potentials that are asymptotic to 
$j=2$ and satisfy this requirement possess less than $10\%$ triplet character at 
$300\;a_0$ and must have binding energies less than $360$~MHz for the classical 
turning point of the wave function to extend out to this region. On the other 
hand, the $0_g^-$ adiabatic potential of the $j=1$ asymptote possesses $20\%$ 
triplet character at $300$~$a_0$ and allows wave functions that correspond to 
binding energies less than $1.8$~GHz to extend out to this region. Consequently, 
the gerade $j=1$ levels are much more likely to be coupled to the metastable 
manifold without suffering significant loss from ionization. 

Applying the same variation in the short-range potentials as was introduced for 
the $j=2$ levels shows some improvement in agreement between the theoretical 
values and the measurements of \cite{Rijn04}, although it is not as large a 
change as might be desired. 

\section{Conclusions}

The availability of the short-range MCSCF and MRCI potentials of \cite{DLGD09} 
enabled us to undertake single-channel and multichannel calculations to study 
the effects of non-adiabatic and Coriolis couplings on the bound states of the 
\mbox{He(2s$\,^{3}$S$_{1}$)+He(2p$\,^{3}$P$_{j}$)} system. The single-channel 
results agree closely with those of \cite{DLGD09} although some discrepancies 
were found that result from the choice of diagonalization grid for the 
single-channel potentials.  Inclusion of non-adiabatic couplings does not affect 
the binding energies very much, but the addition of Coriolis couplings does 
result in some significant and experimentally measurable differences in binding 
energies.  Assignment of theoretical levels to experimental observations, using 
criteria based upon the short-range character of each level and their coupling 
to metastable ground states, reproduces very closely the number of levels 
experimentally observed. The theoretical energies for $j=2$ are consistently 
smaller than the experimental energies by $1.5 - 8\%$, but excellent agreement 
with experiment is obtained after a 1\% increase in the slope of the input 
short-range quintet potentials near the inner classical turning point is 
applied. The same variation also produces an improved matching between the 
theoretical $j=1$ values and measurements of \cite{Rijn04}. 
This suggests the need for improved short-range potentials although such 
adjustments of even the best \textit{ab initio} potentials are not 
uncommon in the theoretical calculation of physical quantities to the accuracy
required in ultracold physics. 

Finally we note that we have chosen to use the same $C_{n}$ parameters for the 
potentials as in \cite{Venturi03} and \cite{DLGD09} rather than the more recent 
coefficients in \cite{Zhang06}, as we wanted to be able to make direct 
comparisons with results obtained in these studies. We have repeated our 
calculations using these new $C_{n}$ values and find that the calculated binding 
energies differ by less than $0.1\%$.  For the unvaried potentials there are 
some differences in the observability criteria and these lead to several 
differences in the assignments.  The addition of the variation to the quintet 
potentials, however, produces exactly the same assignments if either set of 
$C_{n}$ values are used.

\ack
We thank B Deguilhem for providing tabulations of the input MCSCF and MRCI 
potentials and A S Dickinson, J L\'eonard and M van Rijnbach for invaluable 
discussions.
 
\appendix

\section{Decomposition of excited states}\label{app:decomposition}

In order to analyze the singlet, triplet and quintet character of the excited 
bound states we transform from the Hund case (c) basis
$|a\rangle =|\gamma_{1}\gamma_{2}j_{1}j_{2}j\Omega_{j}w \rangle $
to the Hund case (a) basis
$|\alpha \rangle =|\gamma_{1}\gamma_{2}LS\Omega_{L}\Omega_{S}w \rangle $
using
\begin{equation}
\label{cwpB1}
\fl
|\gamma_{1}\gamma_{2}j_{1}j_{2}j\Omega_{j}w\rangle  =  
\sum_{LS\Omega_{L}\Omega_{S}}F_{LS\Omega_{L} \Omega_{S}}^{j_{1}j_{2}j\Omega_{j}} 
|\gamma_{1}\gamma_{2}LS\Omega_{L}\Omega_{S}w\rangle \,,
\end{equation}
where the coupling coefficients 
$F_{LS\Omega_{L}\Omega_{S}}^{j_{1}j_{2}j\Omega_{j}}$ are defined by
\begin{eqnarray}
\label{cwpB2}
\fl
F_{LS\Omega_{L}\Omega_{S}}^{j_{1}j_{2}j\Omega_{j}} =  
[(2S+1)(2L+1)(2j_{1}+1)(2j_{2}+1)]^{\frac{1}{2}}
\nonumber \\  \times
\Cleb{L}{S}{j}{\Omega_L}{\Omega_S}{\Omega_j}
\Ninej{L_1}{L_2}{L}{S_1}{S_2}{S}{j_1}{j_2}{j} \,.
\end{eqnarray}
In (\ref{cwpB2}), $\Cleb{j_1}{j_2}{j}{m_1}{m_2}{m}$ is a Clebsch-Gordan 
coefficient, \mbox{\footnotesize{$\Ninej{a}{b}{c}{d}{e}{f}{g}{h}{i}$}} is the 
Wigner $9-j$ symbol and the implicit set of quantum numbers 
$(\gamma_{1},\gamma_{2})$ has been suppressed.

The singlet, triplet and quintet fractions of the state $G_{a,v}(R)|a \rangle$ 
for $R<20\,a_{0}$, where Penning ionization can take place, are given by
\begin{equation}
\label{cwpB3}
f_{2S+1,v} = \frac{Q_{S,v}}{\sum_{S^\prime} Q_{S^\prime,v}}\,,
\end{equation}
where $S=0,1,2$ and
\begin{equation}
\label{cwpB4}
Q_{S,v} = \sum_{L^\prime S^\prime \Omega_L^\prime \Omega_S^\prime} 
\delta_{SS^\prime} \sum_a F^{j_1 j_2 j \Omega_j}_{L^\prime S^\prime 
\Omega_L^\prime\Omega_S^\prime} \int_0^{20\;a_0} G_{a,v}(R)\;\rmd R \,.
\end{equation}

\section{Radiative coupling}\label{app:coupling}

The approximate coupling between the experimental collision channels and excited 
states is calculated from the matrix element between the metastable state and 
the excited basis states of the interaction $\hat{H}_\mathrm{int} \sim 
\mathbf{\epsilon}_{\lambda}\cdot \mathbf{d}$ for radiation of circular 
polarization $\mathbf{\epsilon}_{\lambda}$ with the molecular dipole operator 
$\mathbf{d}$. The matrix element between basis states of the form (\ref{cwp7}) 
has been derived in \cite{CW09} and is given by
\begin{eqnarray}
\label{cwpA1}
\fl
\langle a^{\prime}| \hat{H}_{\mathrm{int}} | a \rangle  =   (-1)^{\lambda} 
\sqrt{\frac{I}{\epsilon_{0}c}}\; \sqrt{\frac{2J+1}{2J^{\prime}+1}}\; 
F^{j_1^{\prime}j_2^{\prime}j^{\prime}\Omega_j^\prime}_{1,j,-\Omega_j,\Omega_j}
\nonumber \\ \times
\; \sum_{b} \Cleb{J}{1}{J^\prime}{\Omega_j}{b}{\Omega_j^\prime}
\Cleb{J}{1}{J^\prime}{m_J}{\lambda}{m_J^\prime} d_{\mathrm{at}}^{\mathrm{sp}} 
\delta_{w,1 - w^\prime}\,,
\end{eqnarray}
where $\lambda=0,\pm 1$ represents $\pi$ and $\sigma^\pm$ polarization 
respectively in the space-fixed frame, $b$ labels the polarization components in 
the molecular frame, $I$ is the laser intensity and 
$d_{\mathrm{at}}^{\mathrm{sp}}$ is the reduced matrix element of the dipole 
operator between the 2s and 2p atomic states. The basis states $|S m_S l m_l 
\rangle$ that are relevant to experiment, of the system 
\mbox{He(2s$\,^{3}$S$_{1}$)+He(2s$\,^{3}$S$_{1}$)}, are obtained through the 
unitary transformation
\begin{equation}
\label{cwpA2}
\fl
|S m_S l m_l \rangle = \sum_{J m_J j \Omega_j}\delta_{S,j}  (-1)^{j-\Omega_j}
\Cleb{j}{J}{l}{\Omega_j}{-\Omega_j}{0}
\Cleb{S}{l}{J}{m_S}{m_l}{m_J}
|j_1 j_2 j \Omega_j J m_J\rangle\,,
\end{equation}
where for this case $L_1 = L_2 = 0$.

\end{document}